# Image-based measurements of Tafel slopes in aqueous MV/4-HO-TEMPO Flow Batteries

S. Chevalier[1,2,3*], Y. Sasaki[4,5], T. Minami[3,4]

[1] Arts et Métiers Institute of Technology, CNRS, Université de Bordeaux, Bordeaux INP, Institut de Mécanique et d'Ingénierie (I2M), Bâtiment A11, 351 Cours de la Libération, 33405 Talence, France

[2] CNRS, Arts et Métiers Institute of Technology, Université de Bordeaux, Bordeaux INP, Institut de Mécanique et d'Ingénierie (I2M), Bâtiment A11, 351 Cours de la Libération, 33405 Talence, France

[3] LIMMS/CNRS-IIS(UMI2820), The University of Tokyo, 4-6-1 Komaba, Meguro-ku, Tokyo, 153-8505, Japan

[4] Institute of Industrial Science, The University of Tokyo, 4-6-1 Komaba, Meguro-ku, Tokyo, 153-8505

[5] Research Center for Advanced Science and Technology, The University of Tokyo, 4-6-1, Komaba, Meguro-ku, Tokyo,153-8904 Japan

**Abstract**

A total organic aqueous redox flow battery (RFB) employing methyl viologen (MV) electrolyte and 4-hydroxy-2,2,6,6-tetramethylpiperidine-1-oxyl (4-HO-TEMPO) is tested and characterized under microfluidic conditions. The absence of physical membrane and the quasi-two-dimensional energy transfers occurring in this RFB design enables the first direct measurement of Tafel kinetics, charge transfer, and charge transport resistances for both anolyte and catholyte reactants during the RFB operations. The methodology reported in this work combines spectroelectrochemical imaging and analytical modeling of the periodic mass and charge transfer equations. The Tafel kinetics and charge transfers resistances are measured through several RFB geometries and operating conditions without the need of reference electrode nor molar absorptivity calibration, which simplifies the experimental setup, eases the measurements and suppresses the uncertainty related to the electrolyte potential value. Data provided in this work (concentration fields, kinetics properties, electrochemical impedances) quantify the charge transfer in these systems and can serve as reference values for further advanced RFB numerical simulations and design optimization.

**Corresponding author:**

Prof. S. Chevalier
Mechanical Engineering Institute (I2M) of Bordeaux (CNRS UMR 5295)
Arts et Métiers Institute or Technology
Esplanade des Arts et Métiers
33405 Talence Cédex, France
Email: stephane.chevalier@u-bordeaux.fr

## 1. Introduction

Redox flow batteries (RFB) are electrochemical systems capable of converting electricity to large quantities of chemical energy and conversely [1–3]. Most of the reactants used are made of inorganic or organic compounds [4], dissolved mainly in aqueous solvents [5,6]. Few systems also used non-aqueous solvents to take advantages of non-miscible fluids in absence of solid membranes [7]. Vanadium RFB remains the main technology reported in literature [1], but since 10 years, organic RFBs have attracted a lot of attention, particularly because they use environmentally friendly compounds which are generally cheaper to produce than inorganic compounds. Among them, 4-hydroxy-2,2,6,6-tetramethylpiperidine-1-oxyl (4-HO-TEMPO) electrolyte was reported as a promising reactant. Used in combination with 4,4-dimethyl bipyridinium dichloride (methyl viologen, MV), this MV/4-HO-TEMPO RFB has an exceptionally high cell voltage (1 - 1.25 V). Prototypes of the organic RFB reported in literature can be operated at high current densities ranging from 20 to 100 mA/cm$^2$, and deliver stable capacity for 100 cycles with nearly 100% coulombic efficiency [8]. Nevertheless, improvements of these systems are still needed to increase the electrical power density or decrease the pressure drops. Such optimizations rely at first on an accurate knowledge of the kinetics and mass transport properties of the MV/4-HO-TEMPO electrolytes.

The chemical reaction monitoring is still an important challenge to get a better control of these electrochemical technologies [9]. Several methods are reported in literature to measure the electrochemical properties [10], like Tafel kinetics, exchange currents, or current density distributions [11]. Most of them rely on electrochemical characterizations such as electrochemical impedance spectroscopy (EIS) [12], cyclic voltammetry [13], or polarization curve measurements [8] which give a global estimation of these properties. The kinetic rate, $k$ in cm/s, can generally be obtained in situ, see for example Liu et al. [8] , but it is more challenging to measure the Tafel slopes, $b$ in mV. It requires precise measurements of the electrodes overpotential and electrolyte potential which are difficult to perform operando in commercial RFBs. Alternative solutions to answer these challenges were proposed in literature based on microfluidic RFB [14–19]. In most of these works, the microfluidic design of the RFB enable to use liquid electrolyte (so to remove the polymer electrolyte membrane (PEM), but with the drawback of mixing both fluids and limiting the battery cyclability [20–24]). Microfluidic RFBs enable an optical access to measure the concentration fields of the reactants using classical imaging techniques such as visible spectroscopy [25], Raman spectroscopy [26], Fourier transform infrared (FTIR) spectroscopy [27,28] or fluorescence spectroscopy [29]. These imaging techniques, used in combination with electrochemical characterization, are powerful tools to access in operando to RFB kinetic properties and therefore answering the need for advanced characterization systems for electrochemical devices.

The combination of electrochemical and imaging (mainly spectroscopic) techniques has been frequently reported over the last decades with the development of the spectroelectrochemistry [30,31]. These methods enable the simultaneous measurements of electrical parameters, such as current, overpotentials, double layer capacity, charge transfer resistance, and the concentration fields of the reactants. In Garcia et al. [32], spectroelectrochemical imaging was used to characterize the $KMnO_4$ transport and transfer properties in a microfluidic fuel cell. In Matheu et al., it was also used to elucidate the role of some catalysis during reaction [33]. Thus, using similar method, it would be of prime interest propose a direct and easy estimation of the Tafel kinetics, charge transfer and charge transport resistances.

The work reported here precisely answers this question. We propose to combine EIS and visible spectroscopy to create and image the reactant concentration modulation in an operating MV/4-HO-TEMPO RFB. The microfluidic configuration was used to have optical access to the fluid and ensure optimal control of the reactants. Associated with an analytical electrochemical model of the modulated concentration fields, the objective is to measure the Tafel slope and charge transfer resistance of both anolyte and catholyte from the images. We show that this method does not need the three electrodes configuration nor absorptivity coefficient calibration which eases the process and suppresses uncertainties related to the electrode overpotential measurements. Finally, our results are validated by measuring constant Tafel slopes and charge transfer resistances through different operating conditions and charge transport resistances (done by changing the electrolyte concentrations and microchannel geometry).

## 2. Methods

### *1. Spectroelectrochemical measurements in a microfluidic RFB*

An in-house microfuidic RFB was fabricated for imaging purposes. It comprises two Platinum (Pt) electrodes deposited on a Borofloat substrate (2'' wafer). The electrodes are 500 µm wide by 4.5 mm long and they are separated by 1 mm. A polydimethylsiloxane (from Sylgard) T-channel covers the electrode to flow the reactants. Two geometries of microfluidic channel were used: 3 mm wide by 38.5 µm height and 3 mm wide by 59 µm height, ensuring in both cases a large channel aspect ratio (i.e. $l_c/h \gg 1$ here). This particularity reduces the light absorption in the height direction ensuring enough signal to noise ratio (SNR) to use transmission spectroscopy. The top view of the cell is given in Figure 1(b).

Only the charging mode (electrolyze) is studied here. Two aqueous solutions were prepared at the anode using 0.5 M of 4-HO-TEMPO (Tokyo Chemical Industry) plus 1.5 M or 0.75 M of NaCl (Tokyo Chemical Industry) dissolved in DI water. Similarly, at the cathode, two solutions were prepared using 0.5 M of MV (Tokyo Chemical Industry) with 1.5 or 0.75 M of NaCl dissolved in DI water. Both solutions are flowed into the cell using a microfluidic pump (Fluigent Flow EZ), and were degassed before entering the cell using (Systec Mini Vacuum Degassing Chamber) to remove the dissolved oxygen and keep the $MV^{+}$ ions stable after reacting. Electrochemical operating conditions were controlled using a potentiostat (Biologic SP-300) to measure the voltage and the current injected in the cell as well as its impedance using EIS (5 MHz - 100 Hz range). Electrical measurements were performed in a two electrodes configuration (without the use of reference electrode). Oxidation of 4-HO-TEMPO was performed at the working electrode.

Images were collected through an Olympus IX70 inverse microscope. A schematic of our imaging setup is presented in Figure 1(a). The primary light source was a halogen white lamp with collimation adapter placed around 10 cm above the cell. A narrow bandpass filter ($\lambda = 435 \pm 5$ nm from custom made Edmund Optics PN956351) was used to produce a monochromatic blue light passing through the RFB. This wavelength corresponds to the absorption peak of 4-HO-TEMPO [35] and the lowest absorption of MV in the visible domain [36], but since the absorptivity of MV is 100 times higher than 4-HO-TEMPO, the SNR is high enough to image both compounds at the same time using only one wavelength. The light was finally collected through an 8-bit CCD camera (Infinity 2-1R Lumenera). The resulting spatial resolution is 1.86 µm/px leading to a field of view of approximatively 2.5 by 2 mm. It is important to note that in this work, the microscope is used as spectrophotometers to measure the light absorbance and molar concentration of compounds.

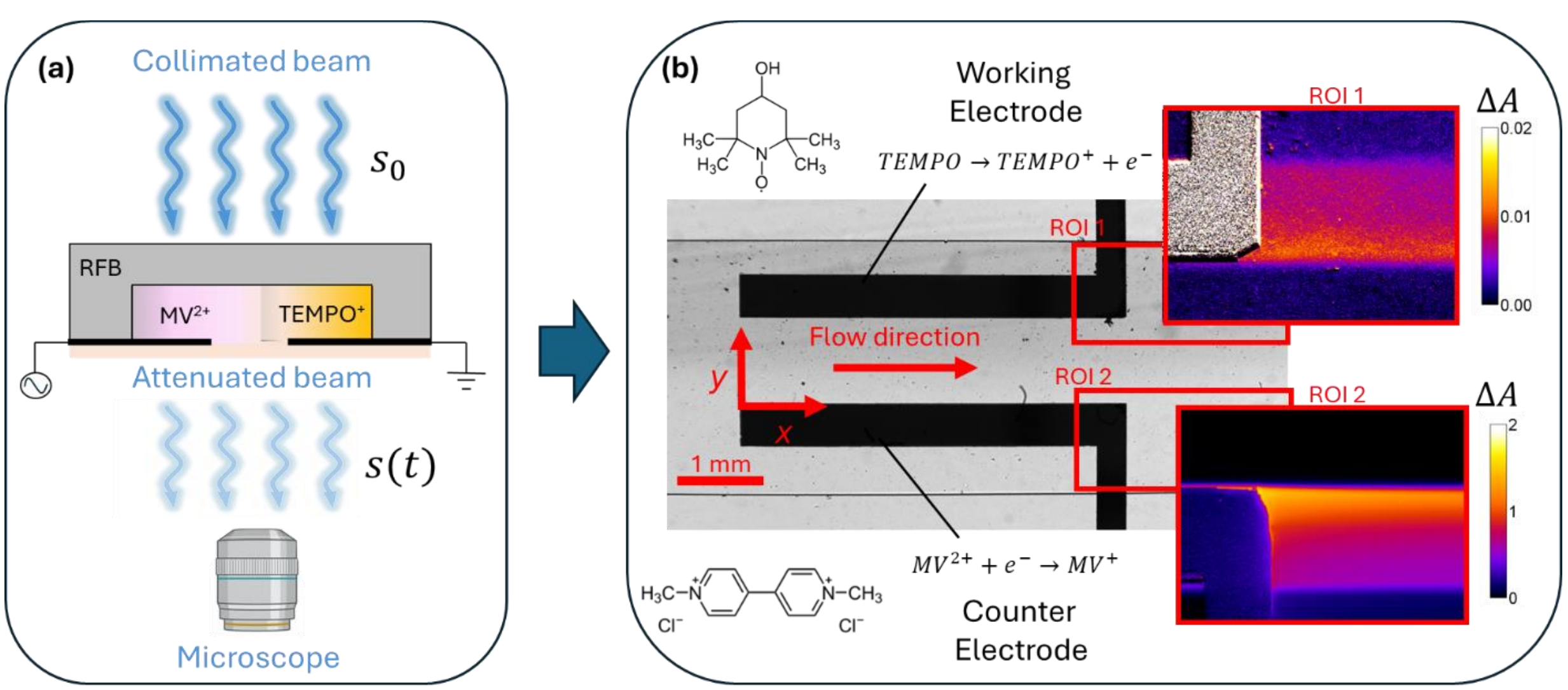

*Figure 1. Schematic of the spectroelectrochemical measurements. (a) The microfluidic RFB with MV/4-HO-TEMPO imaged operando using visible spectroscopy ($\lambda = 435$ nm). $s(t)$ is the signal recorded by the camera proportional to the light intensity. (b) Resulting images of the absorbance variations ΔA observed downstream to the electrodes during 4-HO-TEMPO oxidation and MV reduction of MV (obtained at 1.3 V, $h = 59$ µm, and [NaCl]=1.5 M).*

## 2. *Analytical model of concentration modulation*

To model the 1D behavior of the reactants mass transport in a microfluidic channel, let us consider the average concentration:

$$\bar{c}(x,t) = \frac{1}{l}\int_0^l \frac{1}{h}\int_0^h c(x,y,z,t)\,dydz, \tag{1}$$

where $c$ is the molar concentration, $z$ is the depth (or height) direction, $y$ and $x$ are the width and length direction, respectively, $h$ is the channel height and $l$ is the channel width. The mass transport is done by diffusion and advection, and each phenomena can be characterized by a time scale [19,34] such as the diffusive time $\tau_{dif} \sim l_e^2/(8D)$ and the convective $\tau_{conv} \sim L_e/\bar{v}$, where $L_e$ and $l_e$ are the electrode length and width, respectively, $D$ is the mass diffusivity and $\bar{v}$ is the average fluid velocity. Diffusion can be neglected if $\tau_{diff} \gg \tau_{conv}$ which means that the fluid velocity much ensure $\bar{v} \gg 8DL_e/l_e^2 \sim 0.15$ mm/s (using the electrode dimensions and $D \sim 10^{-3}$ mm$^2$/s). In practice, the velocity will be larger than 2 mm/s, , so the diffusion in x and y direction is negligible, and the 1D concentration field can be described by the following equation [34,37,38]:

$$\frac{\partial \bar{c}}{\partial t} + \bar{v}\frac{\partial \bar{c}}{\partial x} = -\frac{\dot{R}}{h}, \tag{2}$$

where $\bar{v}$ is the average velocity given by $\bar{v} = q_v/S_a$ with $q_v$ the volumetric flow rate and $S_a = hl$ the cross-section area of the microfluidic channel, $h$ is the channel height, assumed small before the width $h \ll l$. The microchannel aspect ratio $h/l \ll 1$ also enables a parabolic velocity profile on top of the electrode (see the solution to Hagen-Poiseuille equation in microchannels given in ref. [22,34]).

The source term, $\dot{R}$, is the chemical consumption or production of reactants on the electrode and it is given by the Butler-Volmer law, which reduces to the Tafel law in the case of large overpotential $\bar{\eta} \gg b$ [39,40] (this assumption is checked in section 3.3). In such cases, one can write the source term as:

$$\dot{R} = \frac{i_0 \bar{c}}{n_e F c^{ref}} \exp\left(\frac{\bar{\eta}}{b}\right) = k\bar{c}, \quad (3)$$

where $i_0$ is the exchange current density, $c^{ref}$ is the reference concentration taken equal as the inlet concentration, $b$ is the Tafel slope, $n_e$ is the number of electrons exchanged, $F$ is the Faraday constant, and $\bar{\eta}$ is the average electrode overpotential. The coefficient $k$ stands for the kinetics reaction rates.

In the case of the spectroelectrochemical measurements, the electrode overpotential is modulated at low frequency $f = \omega/2\pi$, around a steady overpotential, i.e. $\bar{\eta}(t) = \eta_0 + |\delta\eta| \cos(\omega t)$. Such overpotential modulation triggers the concentration modulation at the same frequency. Thus, a convenient way used in literature [37,41,42] to describe such phenomena is to define a complex concentration field made of a steady part, $\bar{c}_0$, and modulated part, $\delta\bar{c}$, [43] as:

$$< \bar{c}(x,t) >= \bar{c}_0(x) + \delta\bar{c}(x,\omega)e^{i\omega t} \quad (4)$$

Injecting this equation in equation 2, leads to the following system in steady state:

$$\frac{d}{dx}\bar{c}_0 = -\frac{k}{h\bar{v}}\bar{c}_0 \quad (5)$$

with $\bar{c}_0(0) = c_{in}$. The modulated part can be written as (see these references for the complete derivation of this equation [21,44,45]):

$$\frac{i\omega}{v}\delta\bar{c} + \frac{d}{dx}\delta\bar{c} = \frac{k}{h\bar{v}}\left(\delta\bar{c} + \frac{\delta\bar{\eta}}{b}\bar{c}_0\right) \quad (6)$$

with $\delta\bar{c}(0) = 0$ as initial conditions. The term $\delta\bar{\eta}$ stands for the average modulation of the electrode overpotential. Solving equations 5 and 6, leads to

$$\bar{c}_0(x) = c_{in} e^{-\frac{k}{h\bar{v}}x}, \tag{7}$$

and

$$\delta\bar{c}(x,\omega) = c_{in}\frac{k\delta\eta}{i\omega hb} e^{-\frac{k}{h\bar{v}}x}\left(1 - e^{-\frac{i\omega}{\bar{v}}x}\right). \tag{8}$$

In order now to simplify equation 8, few extra assumptions can be made. First, only the modulus of the concentration oscillations is considered at the electrode outlet (i.e. $x = L_e$) which leads to:

$$|\delta\bar{c}| = \frac{|\delta\eta|}{b} c_{in}\frac{kL_e}{h\bar{v}} e^{-\frac{kL_e}{h\bar{v}}}|\mathrm{sinc}\,\tilde{\omega}|\,, \tag{9}$$

where sinc is the cardinal sinus function defined as $\sin(x)/x$ and $\tilde{\omega}$ is the dimensionless channel characteristic frequency defined as $\tilde{\omega} = \pi f L_e/\bar{v}$. Note that $|\delta\bar{c}|$ is constant regardless the position downstream to the electrode (but not $\arg(\delta\bar{c})$).

In equation 2, it was assumed that the flow rate in the microfluidic channel is large enough to consider that mass diffusion is negligible. This condition also means that the reactant concentration does not drop significantly at the electrode outlet, as confirmed by the Faraday law in microfluidic $j = n_e F q_v \Delta\bar{c}_0$ [19], which gives for a current $j \sim 0.5$ mA and a flow rate of 5 µl/min $\Delta c \sim 0.15$ M. Compared with $c_{in} = 0.5$ M, the concentration drop is less than 15% in the worst case scenario using the highest current and the lowest flow. Therefore, one can define the variation of molar concentration as $\Delta\bar{c}_0 = c_{in} - \bar{c}_0(L_e) \approx c_{in}\frac{kL_e}{h\bar{v}}$ (using exponential term linearization in equation 8). Then the middle term in equation 9 can be reduced as $c_{in}\frac{kL_e}{h\bar{v}} e^{-\frac{kL_e}{h\bar{v}}} \approx c_{in}\frac{kL_e}{h\bar{v}}(1 - \frac{kL_e}{h\bar{v}}) \approx$ $c_{in}\frac{kL_e}{h\bar{v}} \approx \Delta\bar{c}_0$. Using these linearization, one can end up with a relationship between the modulated concentration, the reactant consumption, and the Tafel kinetics as:

$$\frac{|\delta\bar{c}|}{\Delta\bar{c}_0} = \frac{|\delta\eta|}{b} \ |\operatorname{sinc}\widetilde{\omega}\,| \tag{10}$$

Thus, one can access to $b$ by measuring $|\delta\bar{c}|, \Delta\bar{c}_0$ and $|\delta\eta|$ from spectroelectrochemical measurements. The image processing described in the next section gives the steps needed to measure these values.

## 3. *Image processing*

Each pixel of the camera captured a signal $s(t)$ as schemed in Figure 1(a). Because the voltage and current are modulated around an average value (see Figure 2(a)), it creates a concentration modulation which impacts the light transmission through the RFB. Therefore, the signal recorded by the camera is written as $s(t) = \bar{s} + |\delta s|\cos(\omega t + \arg(\delta s))$ with $\bar{s} = \frac{1}{\tau}\int_0^\tau s(t)dt$ is the average signal and

$$\delta s(\omega) = \int_0^\tau s(t)e^{-2i\pi\omega t}dt, \tag{11}$$

is the Fourier transform of $s(t)$, with $\tau$ equals to an integer number of modulated periods, chosen equal to 10 in this work, such as $\tau = 10/f$.

According to the Beer-Lambert law, the light absorbance, computed from the signal $s(t)$, is proportional to the concentration of RFB reactants, such as $A(t) = \kappa h\bar{c}(t) = \ln(s(t)/s_0)$, where $\kappa$ is the intrinsic molar absorptivity ($M^{-1}.m^{-1}$) of the considered compounds (4-HO-TEMPO or MV), and $s_0$ the light intensity before passing through the RFB (see Figure 1(a)). As said, concentration modulations created by current modulation lead to an absorbance modulation at the same frequency, so we can write that $A(t) = \bar{A} + |\delta A| \cos(\omega t + \varphi_A)$, where $\bar{A}$ is the average absorbance, $|\delta A|$ is the modulus and $\varphi_A$ is to the absorbance modulation phase.

In this work, the average absorbance is not directly measured, but rather the difference between the absorbance measured during Open Circuit Voltage (OCV) and the absorbance measured during RFB operation as:

$$\Delta A = A_{ocv} - \bar{A} = -\ln\left(\frac{\bar{s}}{s_{OCV}}\right) = \kappa h \Delta \bar{c}_0, \tag{12}$$

where $s_{OCV}$ is the pixel signal measured during OCV (before applying the voltage). $\bar{s}$ is the same pixel signal averaged over the entire measurement time after applying the voltage. As shown in equation 12, this quantity is directly proportional to the concentration variation $\Delta \bar{c}_0$ given in equation 10.

The second quantity measured by the camera, also extracted from $A(t)$, is the amplitude of the absorbance modulation defined as

$$|\delta A| \approx \frac{|\delta s|}{\bar{s}} = \kappa h |\delta c|, \tag{13}$$

where $\delta s$ is the Fourier transform of the temporal signal $s(t)$ as defined in equation 11. Equation 13 is obtained from the linearization of Beer-Lambert law using the rule $\ln(1 + \epsilon) \approx \epsilon$.

Finally, the quantities $|\delta A|$ and $\Delta A$, are used to rewrite equation (10) as

$$\frac{|\delta A|}{\Delta A} = \frac{|\delta \bar{c}|}{\Delta \bar{c}_0} = \frac{|\delta \eta|}{b} \, |\operatorname{sinc} \widetilde{\omega}\,|. \tag{14}$$

In this form, it is important to note that it is not needed to calibrate the reactant absorptivities, $\kappa$, to determine their Tafel slope, nor to have the knowledge of the channel height. This is a great asset to our proposed technique as it reduces considerably the experimental process, relies on a limited number of experimental parameters, and can be applied to numerous aqueous solutions by extending the wavelength range (from UV to IR).

## 3. Results

### *1. Spectroelectrochemical measurements*

The MV/4-HO-TEMPO RFB is operated under the microscope using a monochromatic light. Only the charging mode (electrolyze) is studied in this work. In all the experiments, a constant flow rate was used, and a DC + AC voltage was applied (the AC part was kept constant at 50 mV throughout all experiments). It results in a DC + AC current flowing through the electrodes as depicted in Figure 2(a) in the case of 1.3 V DC. Given the low frequency chosen, i.e. 0.25 Hz, both current and voltage are in phase.

The electrodes current modulation also produces both reactants modulation due to the mass charge conservation (Faraday law). The concentration modulations are seen through the signal capture by the camera and presented in Figure 2(b). It is observed that the signal modulation occurs at the same frequency as the current, as expected given the linearity of the Faraday law and Beer-Lambert law. Besides, MV's modulation amplitude is observed 100 times higher than 4-HO-TEMPO meaning that the MV's absorptivity at 435 nm is also 100 times greater than 4-HO-TEMPO [46,47]. Nevertheless, such absorptivity difference of the two compounds is not a problem in this work and absorbance fields were measured simultaneously.

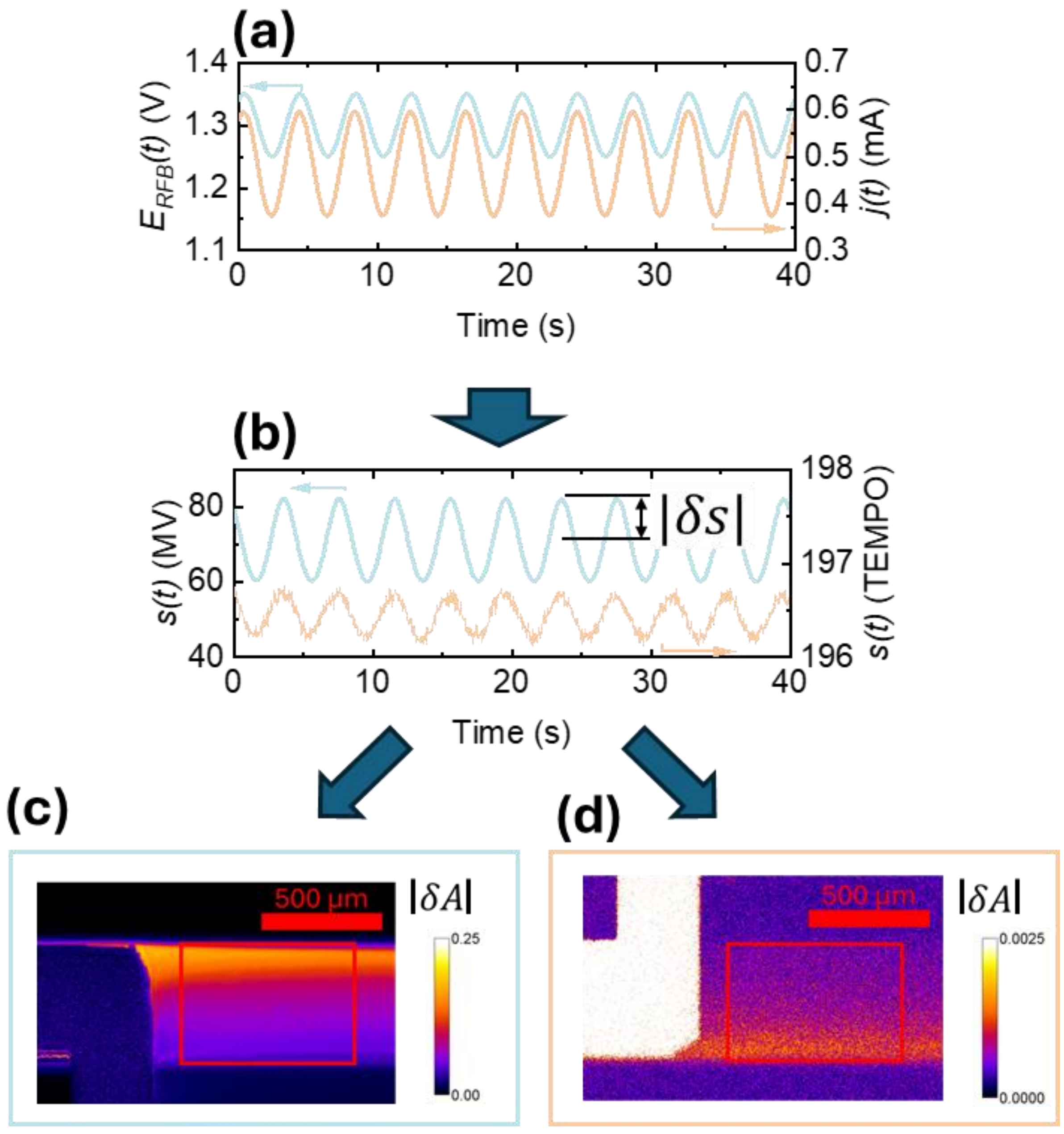

*Figure 2. Signals measured by spectroelectrochemistry at 1.3 V DC,* $h = 59$ *µm, and using 1.5 M of NaCl as electrolyte. (a) Electrical signals: the applied voltage and resulting current at the working electrode during 10 periods. (b) The resulting modulation of the light absorption as it was measured by the camera during the same 10 periods. This signal is an average of a 10 by 10 pixels area downstream to the electrode. (c) and (d) fields of* $|\delta A|$ *obtained after the image processing for MV and 4-HO-TEMPO respectively.*

The resulting modulus of absorbance fields obtained for both MV and 4-HO-TEMPO are given in Figure 2(c) and (d), respectively. These results show the ability of spectroelectrochemistry imaging methods for performing imaging measurements with a noise lower than $|\delta A|_{noise} \sim 10^{-3}$ which means than the signal-to-noise ratio (SNR) is larger than 250 in the case of MV and 2.5 for 4-HO-TEMPO. Such SNR large range could not have been obtained without modulating the RFB voltage and the image processing done in this work [21].

*2. Voltage sweep*

The experimental protocol and image processing described previously were repeated for a range of DC voltages, RFB channel heights and NaCl concentration (1.5 and 0.75 M). All the configurations tested are summarized through the polarization curves in Figure 3. Electrochemical oxidation at the working electrode (4-HO-TEMPO) was observed to start around 1 V, as reported in ref. [48].

The polarization curves measured show that the best RFB performances are obtained with the largest electrolyte concentration and channel heights. The fluid velocity being of the same order of magnitude (2-3 mm/s for each case, see Table 1), so these performance variations are attributed to the change of charge transport resistance. This latter can be estimated as $R_{HF} = l_i/(\sigma h L_e)$ where $l_i$ is the inter-electrode distance (1 mm), $L_e$ is the length of the electrode (4.5 mm), and $\sigma$ is the ionic conductivity (around 0.06-0.12 S/cm depending on [NaCl]). Thus, by decreasing both the NaCl concentration (so the ionic conductivity) and channel height, it results in increasing the charge transport resistance leading to lower RFB performances [8]. This assumption explains well the results on the polarization curve and is also confirmed by measuring the high frequency resistance using EIS (see Table 1).

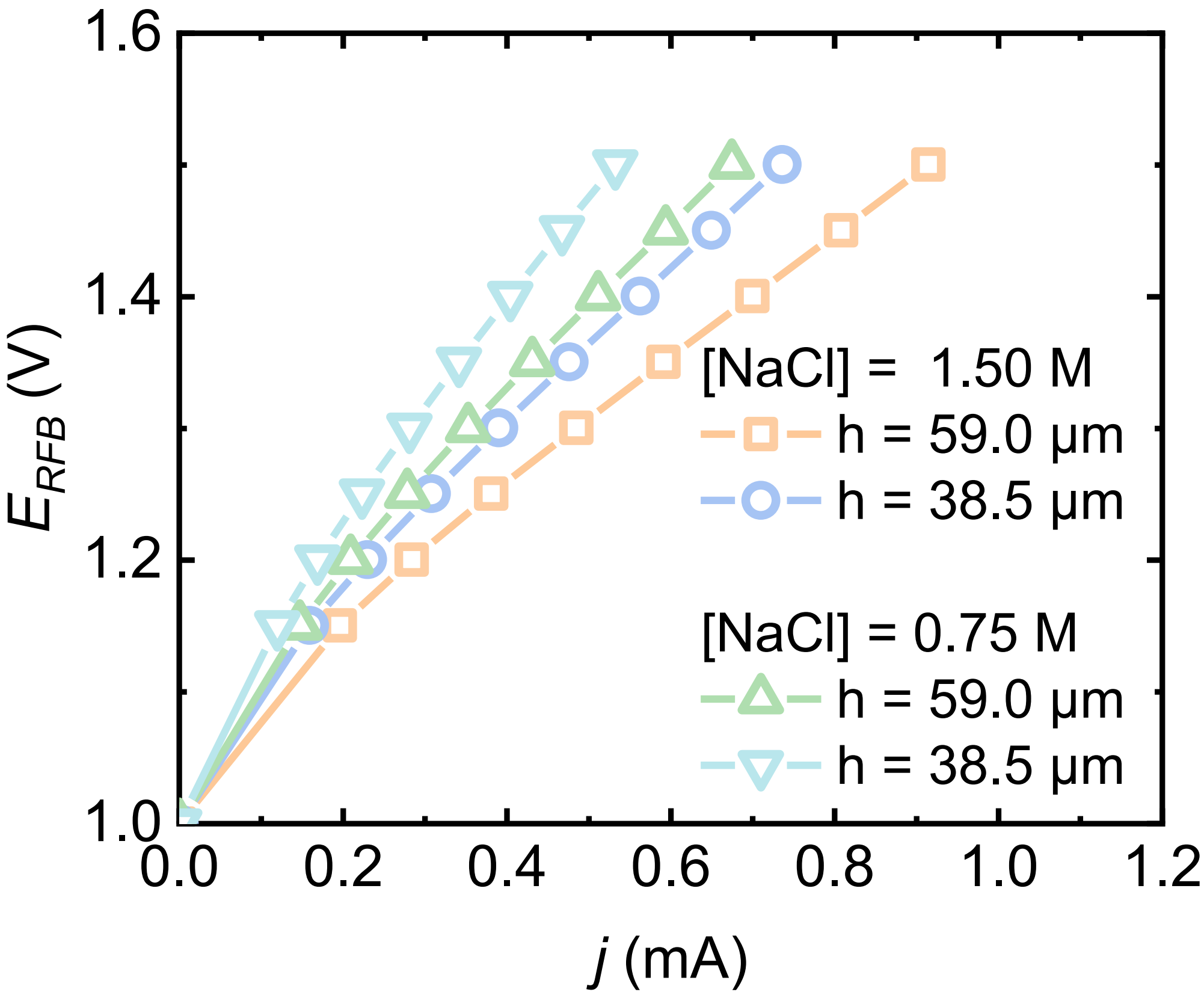

*Figure 3. Polarization curves of the different RFBs tested.*

For each DC voltage measured in Figure 3, the absorbance variations and modulations, $\Delta A$ and $|\delta A|$, respectively, were measured from the images captured by our camera. To ease the visualization, the images were averaged in the x-direction in the red rectangles shown in Figure 2(c) and (d). As observed in Figure 4(a) and (b), $\Delta A$ profiles increase with DC voltage, so the concentrations of $MV^+$ and $TEMPO^+$.

The electrodes are located between the two vertical dash lines in Figure 4(a) and (b). Because of the competition between the charge and mass transport, the electrochemical reaction is not homogeneous on the electrode surface, i.e. the kinetic rates are faster close to the center of the channel. Such behavior was not observed in Garcia et al. as the reaction was limited only by the charge transfer [21]. However, in this study the current is significantly higher, and the charge transport become the limiting factor leading to non-homogeneous kinetics rates as confirmed in literature by several theoretical and experimental studies [45,49,50].

In Figure 4(c) and (d), the absorbance modulation amplitudes are presented. In contrast with the absolute absorbance variations, these quantities are observed constants regardless of the DC voltage. According to equation 10, this result means that the variation of overpotential is balanced by the variation of concentration in the channel. In addition, it is interesting to note that the absorbance modulation amplitudes follow the same profiles as $\Delta A$ in the y-direction. Thus, the charge transport resistance also affects the concentration modulation amplitudes. To the authors' best knowledge, these absorbance profiles are the first obtained operando in a RFB and would be of prime interest to validate RFB numerical model both in steady and periodic state. In the following section, only the average value of $\Delta A$ and $|\delta A|$ are used to measure the Tafel slopes. The profiles are average between the two vertical dashed lines.

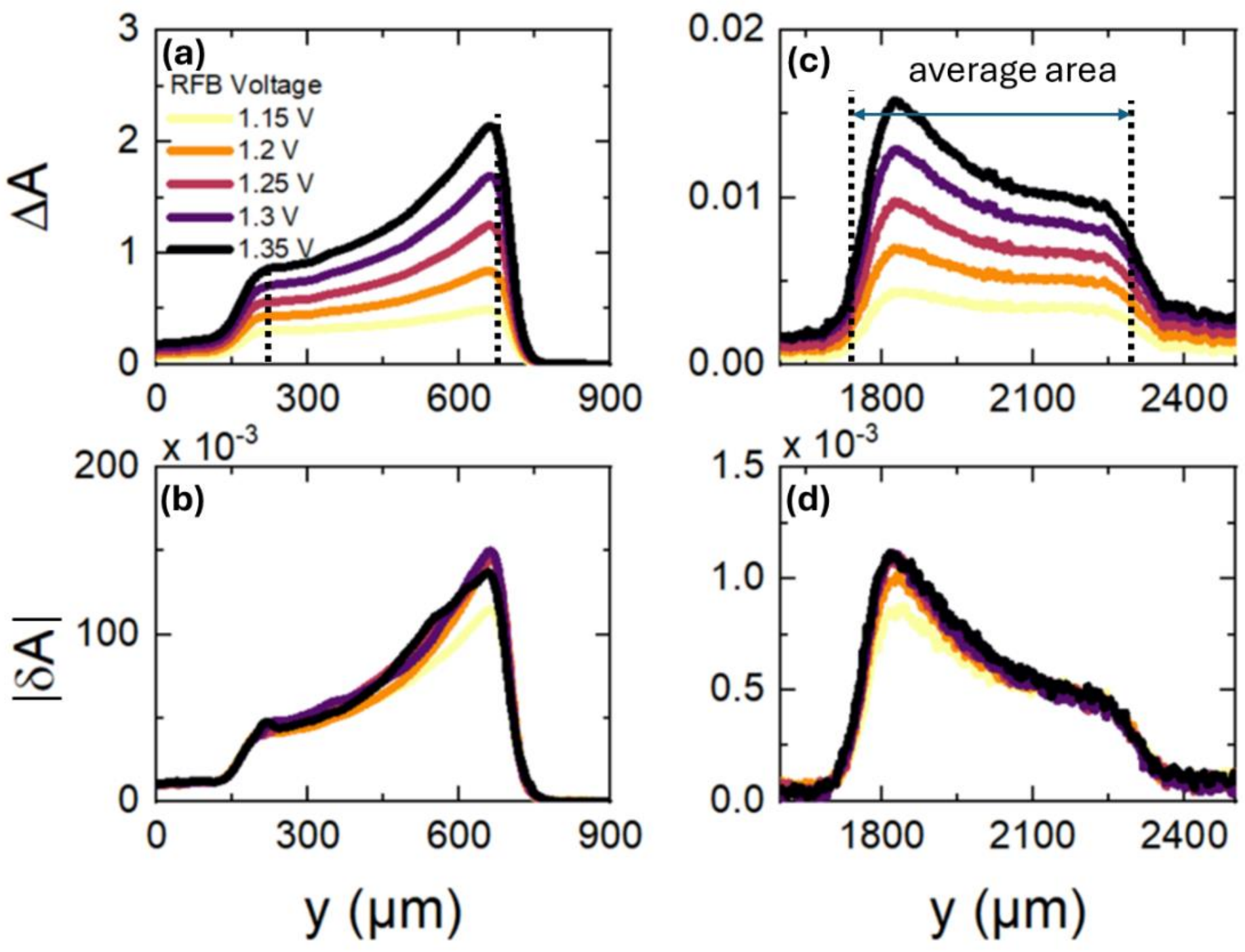

*Figure 4. Absorbance profiles ΔA and absorbance modulation profiles $|\delta A|$ for a range of RFB voltage, obtained with $h = 59$ µm, and using 1.5 M of NaCl as electrolyte. (a) and (b) at the counter electrode outlet (MV), and (c) and (d) at the working electrode outlet (4-HO-TEMPO). The averages were done in the x-direction only in the red rectangles depicted in Figure 2(c) and (d).*

### 3. *Tafel slope measurements*

Moving forward to the Tafel slope measurements, the amplitude of overpotential modulation in equation 10 needs to be estimated . This can be made by considering that the charge transport is 1D in y-direction and does not depend on x, which, once again, is the case using our microfluidic RFB design. This assumption was validated though a 2D numerical simulations (see the supplementary information provided along with this article). In this case, one can write the complex overpotential as:

$$\delta\eta \approx \delta E_{RFB} - R_{tot}\delta j, \tag{15}$$

where $\delta E_{RFB}$ and $\delta j$ are complex numbers describing the modulation of voltage and current (i.e. the Fourier transform of the signals in Figure 2(a)). The resistance $R_{tot}$ describes the total electrical resistance of the RFB comprising the charge transfer resistance $R_{CT}$ and the charge transport resistance $R_{HF}$ obtained using EIS measurements at high frequency ($> 100$ kHz). An equivalent electrical circuit of these resistances is proposed in the insert of Figure 5(c). At low frequency, no phase shift is noticed between $E_{RFB}(t)$ and $j(t)$ (see Figure 2(a)), therefore the imaginary parts of $\delta E_{RFB}$ and $\delta j$ are neglectable and the modulus of equation 15 can be estimated as $|\delta\eta| \approx |\delta E_{RFB}| - R_{tot}|\delta j|$ which finally leads to a convenient relationship to fit both the Tafel slope and the total resistance as

$$\frac{|\delta A|}{\Delta A} = \frac{|\operatorname{sinc}\tilde{\omega}|}{b}\left(|\delta E_{RFB}| - R_{tot}|\delta j|\right). \tag{16}$$

Thus, equation 16 shows that performing several measurements of $|\delta A|/\Delta A$ and $|\delta j|$ for a range of RFB voltage enables us to measure the Tafel slopes from the intercept and the total resistance from the slope using a linear regression. These results are presented in Figure 5(a) and (b) for a range of operating conditions. All the estimated parameters are summarized in Table 1.

In Figure 5(a) and (b), as expected by our model, a linear trend is observed between the modulus of the absorbance and the modulus of the current. This is the case for all the configurations tested, i.e. using different electrolyte concentrations and channel heights, which validates all the assumptions made to derive equation 16. The uncertainty on the estimated parameters measured is 5% in the case of 1.50 M [NaCl] and 2% in the case of 0.75M [NaCl] due to the extended valitidy range in this case (the assumption of low concentration variations holds at higher potential than in the case of 1.5 M [NaCl]). Thus, Tafel slopes of MV and 4-HO-TEMPO are found, in average for all cases, to be around 34 ± 2 mV and 38 ± 2 mV, respectively. They correspond to a transfer coefficient $\alpha = RT/bF$ of 0.68 and 0.76 at room temperature, in excellent agreement with the values measured using rotating electrodes by Janoschka et al. [5]. These values of Tafel slope also validate the choice of Tafel kinetics: by considering at least a minimum overpotential of $\bar{\eta} = 0.1$ V, it leads to $\bar{\eta} \gg 0.038$. In addition, these values are found constant regardless of the electrolyte concentration and the channel geometry which also confirmed that these parameters are only linked to the electrochemical reaction (catalysts and chemicals used).

The total resistance, however, is strongly impacted by the experimental conditions used (see Table 1) from 400 to 700 Ω approximately, but in each case, it is found almost equal for both 4-HO-TEMPO and MV, which is expected since such resistance represents the global charge transfer and charge transport resistances. These observations are also excellent indicators of our measurements robustness. Furthermore, by combining the measurements of the high frequency resistances, $R_{HF}$, done by EIS (Figure 5(d)) and the ones obtained for the total resistance, it is possible to extract the contribution of the charge transfer and transport resistances (according to the model described in the insert of Figure 5(c)). This result is presented in Figure 5(c) where a clear linear trend is observed, and a charge transfer resistance of 104 ± 5 Ω is obtained. Once again, this result agrees well with the assumption that varying the electrolyte concentration and channel height impacts the charge transport resistance but not the Tafel slope and the charge transfer resistance.

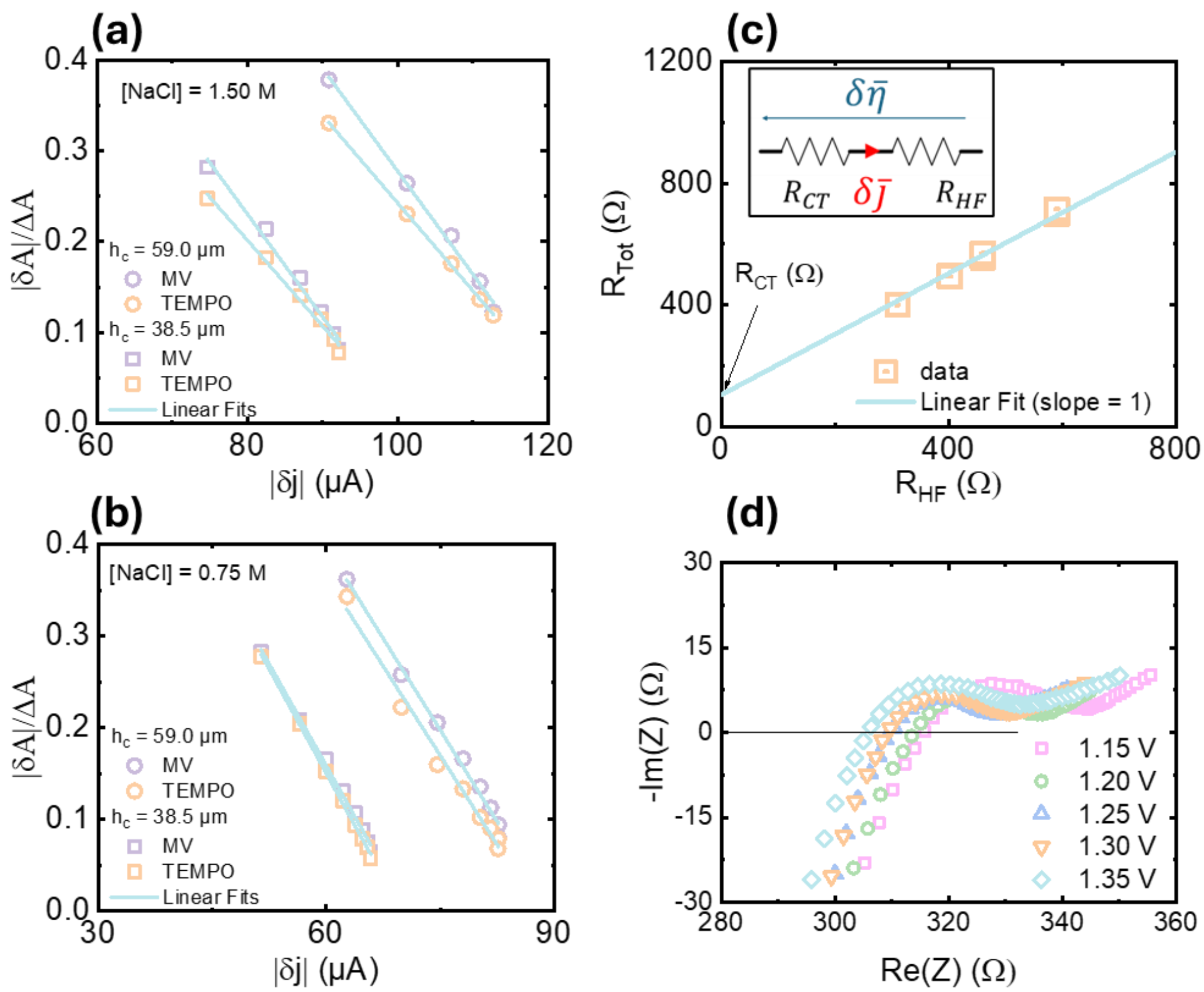

*Figure 5. Tafel slope and charge transfer resistance measurements. (a) and (b) Tafel slope fits from the image-based impedance and current density modulus for 1.50 and 0.75 M of NaCl. (c) Charge transfer resistance estimation. (d) Impedance measurements for* $h = 59$ *μm,[NaCl] = 1.50 M and a range of RFB potentials.*

*Table 1. Summary of the operating parameters and the Tafel slope measurements. The amplitude of voltage modulation was kept constant at* $|\delta E_{RFB}| = 50$ *mV as well as the modulation frequency at 250 mHz throughout all the measurements.*

| **[NaCl] (M)** | 1.50 | | | | 0.75 | | | |
|---|---|---|---|---|---|---|---|---|
| $h$ **(μm)** | 38.5 | | 59 | | 38.5 | | 59 | |
| $q_v$ **(μl/min)** | 10 | | 20 | | 10 | | 20 | |
| $\bar{v}$ **(mm/s)** | 2.89 | | 3.77 | | 2.89 | | 3.77 | |
| $\tilde{\omega}$ **(-)** | 1.22 | | 0.94 | | 1.22 | | 0.94 | |
| $\mathrm{sinc}(\tilde{\omega})$ **(-)** | 0.77 | | 0.86 | | 0.77 | | 0.86 | |
| **Reactants** | MV | 4-HO-TEMPO | MV | 4-HO-TEMPO | MV | 4-HO-TEMPO | MV | 4-HO-TEMPO |

| **intercept (-)** | 1.14 | 0.96 | 1.413 | 1.21 | 1.06 | 1.06 | 1.20 | 1.14 |
|---|---|---|---|---|---|---|---|---|
| **slope (x 1000) (μA$^{-1}$)** | -11.33 | -9.47 | -11.35 | -9.655 | -14.78 | -15.13 | -13.33 | -12.99 |
| $b$ **(mV)** | 34±2 | 40±2 | 30±2 | 36±2 | 36±1 | 36±1 | 36±1 | 38±2 |
| $R_{tot}$ **(Ω)** | 498±25 | 493±25 | 402±20 | 400±20 | 697±14 | 715±14 | 555±11 | 570±11 |
| $R_{HF}$ **(Ω)** | 402±20 | | 310±15 | | 590±15 | | 460±12 | |

## 4. Conclusion

The study of MV/4-HO-TEMPO RFB in microfluidic configuration enabled a thorough characterization of Tafel slopes and charge transfer and transport resistances for a large range of operating conditions and microchannel heights. The use of spectroelectrochemistry based on EIS and visible spectroscopy led to the measurement of image-based impedance. These fields enabled direct access to the RFB electrochemical transfers and absorbance distributions in the channels with an important SNR, from 2.5 to 250. These data are of prime importance to characterize in operando the energy transfers in electrochemical systems and for future numerical model validations.

By varying the charge transport resistance, we were able to measure directly the Tafel slopes of both reactants and the RFB charge transfer resistance. This is, to the authors' best knowledge, the first operando measurements of these essential parameters in an operating RFB. The mathematical framework reported in this work also supports well our results, increasing the confidence and the robustness of the proposed method.

Last but not least, the reported technique was performed in two electrodes configuration without the calibration of reactant absorptivity coefficients. This eases the measurement and suppresses the uncertainty linked to the electrolyte potential and channel geometry, which are usually critical to have a good estimate of concentration fields and electrode overpotentials. In addition, by extending the wavelength range of the setup from UV to IR, it opens the use of our method for characterizing a wide range of organic or inorganic reactants.

**Acknowledgments**

The authors gratefully acknowledge the CNRS and Tokyo University through the Joint PhD Program (SmartBat) for funding this work. Mr Felix Lavanchy is gratefully acknowledged for his assistance during the experiments.

## References

[1] X. Yuan, C. Song, A. Platt, N. Zhao, H. Wang, H. Li, K. Fatih, D. Jang, *A review of all-vanadium redox flow battery durability: Degradation mechanisms and mitigation strategies*, Int. J. Energy Res. **43** (2019) er.4607.

[2] J.W. Lee, M.-A. Goulet, E. Kjeang, *Microfluidic redox battery*, Lab Chip. **13** (2013) 2504.

[3] A.Z. Weber, M.M. Mench, J.P. Meyers, P.N. Ross, J.T. Gostick, Q. Liu, *Redox flow batteries: a review*, J. Appl. Electrochem. **41** (2011) 1137–1164.

[4] A. Clemente, R. Costa-Castelló, *Redox flow batteries: A literature review oriented to automatic control*, Energies. **13** (2020) 1–31.

[5] T. Janoschka, N. Martin, U. Martin, C. Friebe, S. Morgenstern, H. Hiller, M.D. Hager, U.S. Schubert, *An aqueous, polymer-based redox-flow battery using non-corrosive, safe, and low-cost materials*, Nature. **527** (2015) 78–81.

[6] S. Hou, L. Chen, X. Fan, X. Fan, X. Ji, B. Wang, C. Cui, J. Chen, C. Yang, W. Wang, C. Li, C. Wang, *High-energy and low-cost membrane-free chlorine flow battery*, Nat. Commun. **13** (2022) 1–8.

[7] P. Navalpotro, C. Trujillo, I. Montes, C.M.S.S. Neves, J. Palma, M.G. Freire, J.A.P. Coutinho, R. Marcilla, *Critical aspects of membrane-free aqueous battery based on two immiscible neutral electrolytes*, Energy Storage Mater. **26** (2020) 400–407.

[8] T. Liu, X. Wei, Z. Nie, V. Sprenkle, W. Wang, *A Total Organic Aqueous Redox Flow Battery Employing a Low Cost and Sustainable Methyl Viologen Anolyte and 4-HO-TEMPO Catholyte*, Adv. Energy Mater. **6** (2016).

[9] Y. Sasaki, T. Minami, *Organic Field-Effect Transistors for Interfacial Chemistry: Monitoring Reactions on SAMs at the Solid–Liquid Interface*, ACS Appl. Mater. Interfaces. (2025).

[10] Y. Gong, A.L. Clemens, J.T. Davis, C. Orme, N.A. Dudukovic, A.N. Ivanovskaya, R. Akolkar, *Methods—Design Guidelines for Tubular Flow-through Electrodes for Use in Electroanalytical Studies of Redox Reaction Kinetics*, J. Electrochem. Soc. **168** (2021) 043505.

[11] E.A. Stricker, X. Ke, J.S. Wainright, R.R. Unocic, R.F. Savinell, *Current Density Distribution in Electrochemical Cells with Small Cell Heights and Coplanar Thin Electrodes as Used in ec-S/TEM Cell Geometries*, J. Electrochem. Soc. **166** (2019) H126–H134.

[12] M. Cazot, G. Maranzana, J. Dillet, F. Beille, T. Godet-Bar, S. Didierjean, *Symmetric-cell characterization of the redox flow battery system: Application to the detection of degradations*, Electrochim. Acta. **321** (2019) 134705.

[13] L. Zhang, Y. Qian, R. Feng, Y. Ding, X. Zu, C. Zhang, X. Guo, W. Wang, G. Yu, *Reversible redox chemistry in azobenzene-based organic molecules for high-capacity and long-life nonaqueous redox flow batteries*, Nat. Commun. **11** (2020) 1–11.

[14] H. Park, G. Kwon, H. Lee, K. Lee, S.Y. Park, J.E. Kwon, K. Kang, S.J. Kim, *In operando visualization of redox flow battery in membrane-free microfluidic platform*, Proc. Natl.

Acad. Sci. U. S. A. **119** (2022) 1–9.

[15] N. Chaabene, K. Ngo, M. Turmine, V. Vivier, *Ionic liquid redox flow membraneless battery in microfluidic system*, J. Energy Storage. **57** (2023) 106270.

[16] N. Da Mota, D.A. Finkelstein, J.D. Kirtland, C.A. Rodriguez, A.D. Stroock, H.D. Abruña, *Membraneless, room-temperature, direct borohydride/cerium fuel cell with power density of over 0.25 W/cm 2*, J. Am. Chem. Soc. **134** (2012) 6076–6079.

[17] S.A. Mousavi Shaegh, N.-T. Nguyen, S.H. Chan, *A review on membraneless laminar flow-based fuel cells*, Int. J. Hydrogen Energy. **36** (2011) 5675–5694.

[18] Y. Li, W. Van Roy, P.M. Vereecken, L. Lagae, *Effects of laminar flow within a versatile microfluidic chip for in-situ electrode characterization and fuel cells*, Microelectron. Eng. **181** (2017) 47–54.

[19] O.A. Ibrahim, M. Navarro-Segarra, P. Sadeghi, N. Sabaté, J.P. Esquivel, E. Kjeang, *Microfluidics for Electrochemical Energy Conversion*, Chem. Rev. **122** (2022) 7236–7266.

[20] A. Bazylak, D. Sinton, N. Djilali, *Improved fuel utilization in microfluidic fuel cells: A computational study*, J. Power Sources. **143** (2005) 57–66.

[21] M. Garcia, A. Sommier, D. Michau, J.-C. Batsale, S. Chevalier, *Operando Spectroelectrochemical Imaging of Concentration Fields and Tafel Kinetics in Microfluidic Electrochemical Devices*, Anal. Chem. **96** (2024) 16487–16492.

[22] S. Chevalier, *Semianalytical modeling of the mass transfer in microfluidic electrochemical chips*, Phys. Rev. E. **104** (2021) 035110.

[23] P. Navalpotro, C.S. Santos, M.L. Alcantara, V. Muñoz-Perales, S.E. Ibañez, A. Martínez-Bejarano, N. Jiyane, C.M.S.S. Neves, R. Rubio-Presa, T. Quast, W. Schuhmann, J.A.P. Coutinho, R. Marcilla, *Enhancing the Stability of Aqueous Membrane-Free Flow Batteries: Insights into Interphase Processes*, Angew. Chemie - Int. Ed. **64** (2025).

[24] M.J. Torres, J. Hervas-Ortega, B. Oraá-Poblete, A.B. de Quirós, A.A. Maurice, D. Perez-Antolin, A.E. Quintero, *Membraneless Micro Redox Flow Battery: From Vanadium to Alkaline Quinone*, Batter. Supercaps. **202400331** (2024) 1–7.

[25] J.K. Utterback, A.J. King, L. Belman-Wells, D.M. Larson, L.M. Hamerlynck, A.Z. Weber, N.S. Ginsberg, *Operando Label-Free Optical Imaging of Solution-Phase Ion Transport and Electrochemistry*, ACS Energy Lett. **8** (2023) 1785–1792.

[26] F. Sarrazin, J.-B. Salmon, D. Talaga, L. Servant, *Chemical Reaction Imaging within Microfluidic Devices Using Confocal Raman Spectroscopy: The Case of Water and Deuterium Oxide as a Model System*, Anal. Chem. **80** (2008) 1689–1695.

[27] K. Krause, C. Palka, M. Garcia, A. Erriguible, S. Glockner, J.-L. Battaglia, S. Chevalier, *Heat and mass transfer visualisation of an exothermic acid-base microfluidic reactor using infrared thermospectroscopy*, Quant. Infrared Thermogr. J. **00** (2024) 1–18.

[28] S. Chevalier, J.-N. Tourvieille, A. Sommier, C. Pradère, *Infrared thermospectroscopic imaging of heat and mass transfers in laminar microfluidic reactive flows*, Chem. Eng. J. Adv. **8** (2021) 100166.

[29] R. Kuriyama, T. Nakagawa, K. Tatsumi, K. Nakabe, *Two-dimensional fluid viscosity measurement in microchannel flow using fluorescence polarization imaging*, Meas. Sci.

Technol. **32** (2021) 095402.

[30] J. Garoz-Ruiz, J.V. Perales-Rondon, A. Heras, A. Colina, *Spectroelectrochemical Sensing: Current Trends and Challenges*, Electroanalysis. **31** (2019) 1254–1278.

[31] J.R. Fish, S.G. Swarts, M.D. Sevilla, T. Malinski, *Electrochemistry and spectroelectrochemistry of nitroxyl free radicals*, J. Phys. Chem. **92** (1988) 3745–3751.

[32] M. Garcia, A. Sommier, D. Michau, G. Clisson, J.-C. Batsale, S. Chevalier, *Imaging concentration fields in microfluidic fuel cells as a mass transfer characterization platform*, Electrochim. Acta. **460** (2023) 142489.

[33] R. Matheu, M.Z. Ertem, M. Pipelier, J. Lebreton, D. Dubreuil, J. Benet-Buchholz, X. Sala, A. Tessier, A. Llobet, *The Role of Seven-Coordination in Ru-Catalyzed Water Oxidation*, ACS Catal. **8** (2018) 2039–2048.

[34] H. Bruus, *Theoretical microfluidics*, Oxford university press, 2008.

[35] K. Żamojć, M. Zdrowowicz, W. Wiczk, D. Jacewicz, L. Chmurzyński, *Dihydroxycoumarins as highly selective fluorescent probes for the fast detection of 4-hydroxy-TEMPO in aqueous solution*, RSC Adv. **5** (2015) 63807–63812.

[36] N. Aristov, A. Habekost, *Electrochromism of Methylviologen (Paraquat)*, World J. Chem. Educ. **3** (2015) 82–86.

[37] S. Chevalier, J.-C. Olivier, C. Josset, B. Auvity, *Polymer electrolyte membrane fuel cell operating in stoichiometric regime*, J. Power Sources. **440** (2019) 227100.

[38] S. Chevalier, M. Garcia, A. Sommier, J.-C. Batsale, *Semianalytical mass transfer impedance model in microfluidic electrochemical chips*, (2022).

[39] E.J.F. Dickinson, A.J. Wain, *The Butler-Volmer equation in electrochemical theory: Origins, value, and practical application*, J. Electroanal. Chem. **872** (2020) 114145.

[40] A. Taqieddin, M.R. Allshouse, A.N. Alshawabkeh, *Editors' Choice—Critical Review—Mathematical Formulations of Electrochemically Gas-Evolving Systems*, J. Electrochem. Soc. **165** (2018) E694–E711.

[41] A. Kulikovsky, O. Shamardina, *A Model for PEM Fuel Cell Impedance: Oxygen Flow in the Channel Triggers Spatial and Frequency Oscillations of the Local Impedance*, J. Electrochem. Soc. **162** (2015) F1068–F1077.

[42] M. Eikerling, A.A. Kornyshev, *Electrochemical impedance of the cathode catalyst layer in polymer electrolyte fuel cells*, J. Electroanal. Chem. **475** (1999) 107–123.

[43] G. Maranzana, J. Mainka, O. Lottin, J. Dillet, A. Lamibrac, A. Thomas, S. Didierjean, *A proton exchange membrane fuel cell impedance model taking into account convection along the air channel: On the bias between the low frequency limit of the impedance and the slope of the polarization curve*, Electrochim. Acta. **83** (2012) 13–27.

[44] J. Mainka, G. Maranzana, J. Dillet, S. Didierjean, O. Lottin, *Effect of Oxygen Depletion Along the Air Channel of a PEMFC on the Warburg Diffusion Impedance*, J. Electrochem. Soc. **157** (2010) B1561.

[45] A.A. Kulikovsky, *Exact low–current analytical solution for impedance of the cathode catalyst layer in a PEM fuel cell*, Electrochim. Acta. **147** (2014) 773–777.

[46] T.N. Das, T.K. Ghanty, H. Pal, *Reactions of methyl viologen dication (MV2+) with H atoms in aqueous solution: Mechanism derived from pulse radiolysis measurements and ab initio MO calculations*, J. Phys. Chem. A. **107** (2003) 5998–6006.

[47] J.B. Gerken, S.S. Stahl, *High-potential electrocatalytic O2 reduction with nitroxyl/NOx mediators: Implications for fuel cells and aerobic oxidation catalysis*, ACS Cent. Sci. **1** (2015) 234–243.

[48] S. Hu, L. Wang, X. Yuan, Z. Xiang, M. Huang, P. Luo, Y. Liu, Z. Fu, Z. Liang, *Viologen-Decorated TEMPO for Neutral Aqueous Organic Redox Flow Batteries*, Energy Mater. Adv. **2021** (2021) 1–8.

[49] S. Chevalier, B. Auvity, J.C. Olivier, C. Josset, D. Trichet, M. Machmoum, *Detection of Cells State-of-Health in PEM Fuel Cell Stack Using EIS Measurements Coupled with Multiphysics Modeling*, Fuel Cells. **14** (2014) 416–429.

[50] J. Mainka, G. Maranzana, J. Dillet, S. Didierjean, O. Lottin, *On the estimation of high frequency parameters of Proton Exchange Membrane Fuel Cells via Electrochemical Impedance Spectroscopy*, J. Power Sources. **253** (2014) 381–391.